\documentclass{bmcart}
\usepackage{graphicx}
\usepackage{fontenc}
\usepackage[utf8]{inputenc}
\usepackage{microtype}
\usepackage{url}
\usepackage{cite}
\usepackage{booktabs}
\usepackage[colorlinks = true,
            urlcolor  = black, 
            linkcolor = black,
            citecolor = black, 
            bookmarks = true]{hyperref}
\usepackage{booktabs} 
\usepackage{float}
\begin{document}

\begin{frontmatter}

\begin{fmbox}
\dochead{Research}

\title{Analysis of co-authorship networks among Brazilian graduate programs in computer science}
%
%
%
%
%
%

\author[
   addressref={aff1},                   
   email={alexsilva@alunos.utfpr.edu.br, ORCID 0000-0002-2157-8050}
]{\inits{AJNS}\fnm{Alex Junior Nunes da} \snm{Silva}}
\author[
   addressref={aff1, aff3},
   email={matheus.m.breve@gmail.com}
]{\inits{MMB}\fnm{Matheus Montanini} \snm{Breve}}
\author[
   addressref={aff2},                   
   email={jesus.mena@ufabc.edu.br}   
]{\inits{JPMC}\fnm{Jesús P.} \snm{Mena-Chalco}}
\author[
   addressref={aff1},                   
   corref={aff1},                       
   email={fabricio@utfpr.edu.br, ORCID 0000-0002-8786-3313}   
]{\inits{FML}\fnm{Fabrício M.} \snm{Lopes}}


\address[id=aff1]{
  \orgname{Universidade Tecnológica Federal do Paraná (UTFPR)}, 
  \street{Av. Alberto Carazzai, 1640},                     %
  \postcode{86300-000}                                
  \city{Cornélio Procópio},                              
  \cny{Brazil}                                    
}
\address[id=aff2]{%
  \orgname{Universidade Federal do ABC (UFABC)},
  \street{Alameda da Universidade},
  \postcode{09606-045}
  \city{São Bernardo do Campo},
  \cny{Brazil}
}
\address[id=aff3]{%
  \orgname{Instituto Politecnico de Braganca (IPB)},
  \street{Campus de Santa Apolónia},
  \postcode{5300-253}
  \city{Bragança},
  \cny{Portugal}
}


\begin{artnotes}
\end{artnotes}

\end{fmbox}


\begin{abstractbox}

\begin{abstract} 
The growth and popularization of platforms on scientific production have been the subject of several studies, producing relevant analyses of co-authorship behavior among groups of researchers. Researchers and their scientific productions can be analyzed as co-authorship social networks, so researchers are linked through common publications. In this context, co-authoring networks can be analyzed to find patterns that can describe or characterize them. This work presents the analysis and characterization of co-authorship networks of academic Brazilian graduate programs in computer science. To this end, data from the curricula of Brazilian researchers were collected and modeled as co-authoring networks among the graduate programs that researchers participate in. Each network topology was analyzed regarding complex network measurements and three qualitative indices that evaluate the publication's quality. In addition, the co-authorship networks of the graduate programs were characterized in relation to the evaluation received by CAPES, which attributes a qualitative grade to the graduate programs in Brazil. The results indicate some of the most relevant topological measures for the program's characterization and evaluate at different qualitative rates and indicate a pattern of the graduate programs best evaluated by CAPES.

\end{abstract}


\begin{keyword}
\kwd{bibliometrics}
\kwd{complex networks}
\kwd{topological measurements}
\kwd{pattern recognition}
\kwd{seniority index}
\end{keyword}


\end{abstractbox}
%

\end{frontmatter}


%
%
%
%
%
%
\section*{Introduction}


A large volume of data is generated every day in the most varied areas and contexts. In particular, the production of scientific articles linked to researchers and graduate programs. Different works have tried to analyze academic-related networks, such as co-authorship networks, often constituted by nodes and links, representing researchers and papers written in mutual collaboration, respectively \cite{Savic2019}. Digiampietri \textit{et al}. analyzed and characterized Brazilian Computer Science graduate programs using data available on the Lattes Platform \cite{digiampietri2014}. Mena-Chalco \textit{et al}. described co-authorship networks between Brazilian researchers. They expanded Digiampietri's data set and analyzed all available major knowledge areas to understand the network structures and dynamics, based on data from the Lattes Platform \cite{scriptLattes2009,menachalco2014}. Other authors performed a similar analysis using a different data source, for example, Glänzel's paper in which the Web of Science platform was used \cite{glanzel2006}. These analyses are not restricted to Brazilian networks or computer science researchers, with authors exploring co-authorship networks in other countries such as Turkey \cite{cavusglu2013} or within other areas, such as analyzing relationships among mathematicians \cite{cerinsek2015}.

The works aforementioned do not address the evaluation considering the topological network measurements and a quality indicator. Lopes \textit{et al.} analyzed the relationship between a quality assessment and internal academic collaborations among researchers in Brazilian Computer Science graduate programs, adopting the \textit{Digital Bibliography \& Library Project} (DBLP) database instead of the Lattes Platform \cite{lopes2011ranking}. Linden \textit{et al.} compared high-quality Brazilian graduate programs to international programs of excellence based on different universities and citations rankings, for example, the Shanghai and The Guardian university rankings \cite{linden2017}. 

The evaluation of Brazilian graduate programs, which offer a master’s degree and the doctoral degree, is performed by CAPES. The Coordination of Superior Level Staff Improvement (in portuguese Coordenação de Aperfeiçoamento de Pessoal de Nível Superior or CAPES) \cite{nota-capes} is a governmental institution of the Ministry of Education in Brazil, which has among its activities the evaluation of graduate programs and investments in the formation of high-level resources in the country and abroad. The assessment is carried out in periods, and until 2012 the periods were 3 years; however, due to CAPES changes, from 2013 the assessment period is every 4 years. At each evaluation period, the Brazilian graduate programs receive a qualitative grade from 1 to 7, called  \textit{Nota CAPES} (CAPES grade), being 1 the worst evaluation and 7 the best evaluation. The rate received leads to several consequences for the programs, such as programs with grades 1 and 2 are not recommended by CAPES, programs with grades 3 may offer only masters courses, programs with grades 4 or higher may offer masters and Ph.D. courses, programs with grades 5 or higher may participate in some government research funding initiatives, while others do not, to name a few. Based on these premises, we decided to investigate how academic co-authorships within Brazilian Computer Science graduate programs are related to the qualitative CAPES grade.


%
%
%
%

In this context, this work presents an approach of co-authorship network analysis of graduate programs to characterize and identify topological patterns that can explain the quality grade received from CAPES, considering three evaluation periods. 
For this, we adopted some complex network and vulnerability measurements. We adopted the \textit{Lattes Platform} \cite{lattes} and the \textit{Sucupira Platform} \cite{sucupira}, as data source about Brazilian researchers, their publications, and researcher's affiliation to graduate programs, with these data, we model the graduate co-authorship networks. The concept of academic collaboration can be broad. Researchers can collaborate to produce technical artifacts (software, patents, etc.), with research projects, teaching projects, among other forms; however, this project's scope analyzes as collaboration only publications between researchers.

The achieved results can be of great relevance for the coordination of graduate programs, given the factors that differentiate the best evaluated from the least evaluated programs. That can bring the improvement of the indicators and, consequently, improve the evaluation of the graduate programs.

\subsection*{Metrics}

Complex network theory has been successfully applied in many areas, particularly within representation of networks of different types, such as biological systems \cite{lopes2011b,lopes2014a,ito2018}, computer vision \cite{backes2013,lima2015,piotto2016,lima2019}, the electric power grid \cite{albert2004structural}, the Internet \cite{maslov2004detection}, subway systems \cite{angeloudis2006large}, and neural networks \cite{kotter2003network}, to cite but a few. Another area in which they are also applied is the representation of friendship networks or collaborations between individuals.

Numerical quantifiers are needed to analyze these networks. For that purpose, we adopted network measurements. These are quantifiers that are related to a specific network behavior as a whole. Boccaletti \textit{et al.} \cite{boccaletti2006complex} and Costa \textit{et al.} \cite{costa2007characterization} discuss the many available measurements that can be extracted from a network.

In this context, we adopted 42 network measurements to explore a broad range of possible behavior. Among them, measures of complex network structure (topological), vulnerability measures analysis, and measures that evaluate the position in which the researcher's name is among the publication authors. 


\subsection*{Related works}


The authors in Digiampietri \textit{et al.} \cite{digiampietri2014} present different ways of visualizing the performance of programs, rank them according to each perspective raised and combine them; they also show the correlation between the different metrics, discuss how the programs interact. The authors evaluate the programs in two different ways: the first concerning the productivity of a program based on bibliometric indexes (SJR, JCR impact factor, and Qualis), showing the number of citations, g-index, an h-index of the Microsoft Academic Search and Google Scholar machines; the second form of evaluation takes into account the academic social networks, where graphs are created in which the nodes are the programs and the edges, the interaction between these programs, in that evaluation the considered period is between 2004 and 2009 and the data are extracted of Curriculum Lattes. In this work, it is possible to verify through the measures of productivity the program's evolution, and it is also observed that the rankings of productivity can have considerable variation depending on the used feature. With the work's results, the authors conclude that the best programs located in the network's topology are the most productive.

Two of Newman's works  \cite{Newman2001b} and \cite{Newman2001} in which it is analyzed publications between 1995 and 1999 in the areas of physics, biomedical research, and computer science. Social networks were created to examine each of these areas, including some metrics as the average of researchers per publication, the average number of publications by researchers, clustering coefficient, presence and size of the giant component, distance measurements between researchers, and even measures of centrality in the networks. In analyzing the generated data in these works, it is possible to observe that the distributions of author quantities by papers and works by authors follow a mathematical power-law. It was also noted that all analyzed networks had one component connected to most network elements (Giant component). It was also observed that the network structures have differences between them. Another important notice is that the analyzed networks have small distances between the authors' pairs, which classify them as small-world networks. Besides, most of the authors, most of the paths between them and other researchers, pass only one or two other researchers.

In Bordin \cite{bordin2014}, the authors analyzed the bibliographic production through the co-authorship networks of the Engineering and Knowledge Management program of the Federal University of Santa Catarina in Brazil between 2005 and 2012, they used a metric that is the average number of students, authors by bibliographic production where the total number of researchers (nodes) of a program is divided by the total number of publications (edges), and the more productive a program, the higher this indicator will be. Thus, the authors concluded that this index remained unchanged, stable over the period analyzed. With some complex networks measurements analyzed, the authors were able to classify the authors with higher degree centrality, therefore, the most influential in each program and with the results of the work, they also concluded that in the first year of the program, the percentage of individual publications was the highest, because the program was new and there was not a large amount of co-authored work, despite analyzing the measures the authors did not make any qualitative analysis of the publications and did not compare with the CAPES grade or any other evaluation.

\section*{Materials and Methods}

\begin{figure}[h!]
\centering 
\includegraphics[width=0.95\textwidth]{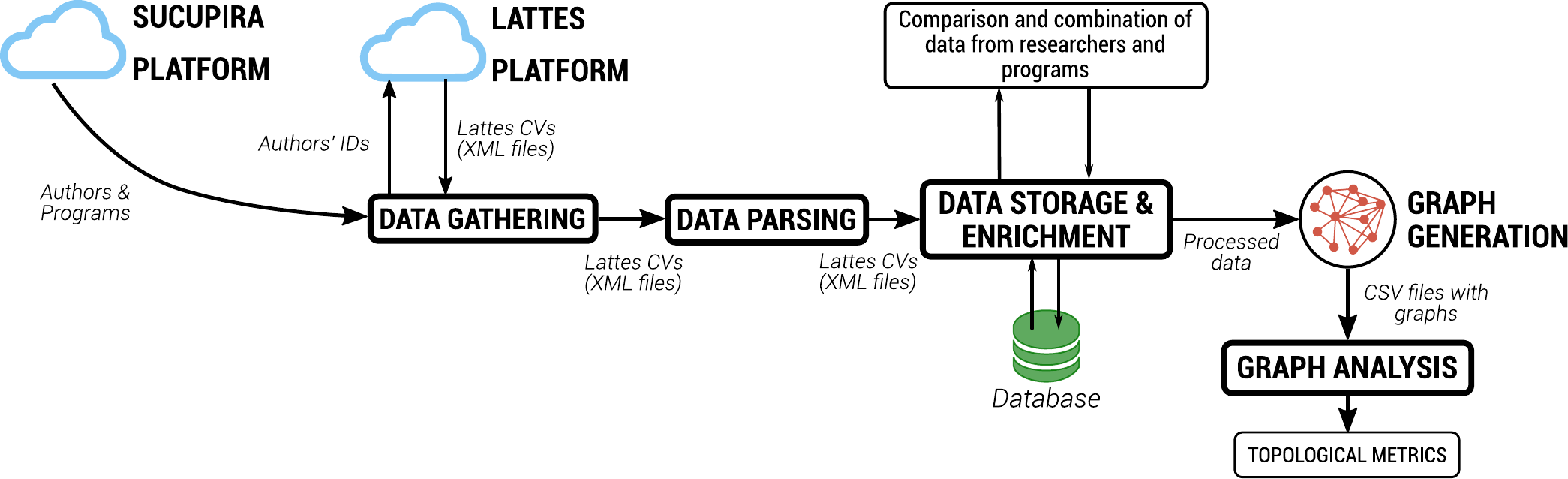}
\caption{\csentence{Process Flow.} Clouds indicate internet data access. Blocks in bold represent processes. Artifacts generated by each process are described next to the arrows that indicate the directions of the flow.}
\label{fig:Process Flow2}
\end{figure}

This subsection presents the materials and methods used to extract and process data to create the graduate co-authorship networks. These networks are then analyzed through complex networks measurements. Fig.~\ref{fig:Process Flow2} presents an overview of the proposed approach.

\subsubsection*{Data sources}

The Lattes Platform \cite{lattes} is a Brazilian online platform in which researchers can create their academic résumés and list their publications. This platform is used nationwide as a decision factor for hiring university staff, distributing federal financial support for research and university scoring, among other uses. Researchers are also evaluated based on the literary production listed in their Lattes résumés, a determinant factor to obtain research grants, for example. As a consequence, governmental institutions encourage researchers to keep their résumés updated and complete. The Lattes platform counts with more than 6 million online résumés. It provides a significant amount of reliable data extracted and analyzed to determine the main features that distinguish low-ranked and high-ranked universities based on their researchers' academic performances. The Lattes résumés is so relevant that some of the data present in Sucupira comes from it.
The Lattes Résumés of 1,644 researchers, affiliated with Brazilian Computer Science graduate programs were extracted from the Lattes Platform in June 2019. The data acquisition process is further detailed in the section ``Data gathering.''

The quality of Brazilian graduate programs is assessed since 1998 by CAPES. From 1998 to 2014, the quality assessment was released every 3 years, changing to a 4-year interval between examinations after 2014. Graduate programs are analyzed and granted a numerical score, denominated ``\textit{Nota CAPES}'' (which is the Portuguese term for CAPES grade), which ranges from 1, the minimum possible score, to~7, attributed to institutions of excellence. The results of the evaluation are publicly accessible on the CAPES website. Institutions evaluated with ratings 1 or 2 have their authorizations to operate revoked. Institutions offering only master's degrees can obtain a minimum score of 3, while those offering Ph.D. degrees as well are eligible to a minimum score of 4. Scores above 5 are granted to institutions with excellent academic performance. Still, programs with a score of 7 are programs that can be compared with the best world institutions of higher education. The CAPES assessment is carried out with internal analyzes with information systems and professionals who, through studies, reach the CAPES score; more detailed technical data on how CAPES performs this assessment can be accessed through the area document and the evaluation form \cite{CAPES2020}\cite{CAPES2019}.


\subsubsection*{Data gathering}

The \textit{Lattes Platform}, in addition to data on the professional performance of users (hiring in public and private companies), also provides information on researchers and their publications. However, their academic affiliation is not always made clear,  as well as other diverse data, since everything is manually defined by users. To find out the researchers that belong to a given institution and graduate program, 89,255 Brazilian researchers' records were gathered from the  \textit{Sucupira Platform}. This platform contains relevant data of graduate programs, such as, lists containing the employed researchers in Brazilian universities. From this platform, the records of 89,255 Brazilian researchers were extracted.

The data from the \textit{Sucupira Platform} allowed the creation of a list of graduate programs and their respective researchers. This list was filtered to include only researchers in Computer Science and academic-oriented graduate programs, thus leaving out all other researchers from other areas or professional-oriented programs. We get a list of 1,644 Computer Science researchers' full names and their affiliations with this process.


The previous process resulted in a list with all the researcher's names in Computer Science graduate programs. With this list, we download each researcher's Curriculum Lattes from the Lattes Platform in an XML format. We build transformations to convert the semi-structured data (XML) into structured data (DBMS). For this, we create a PostgreSQL-based database management system. 
Within each Curriculum Lattes, are the personal and academic information of each researcher. Also present in the CV is information about their publications. This information has to be manually added by the user in his/her CV via the Lattes website. 
We used all these publications to populate a DBMS table, however, only publications classified as ``\textit{work in events}'' and with ``\textit{COMPLETE}'' type as well as articles (conference publications) with the ``\textit{COMPLETE}'' nature are considered. 

We made several combinations of data to keep the database cohesive and, using Sucupira's information; we linked researchers to programs and institutions. As previously stated, the programs were evaluated every three years, but as of 2014, by CAPES decision, the period changed to four years. In this way, CAPES evaluates each of the programs and assigns them a score for that specific time interval. We inserted these evaluations in the database, so, for further analysis, they can be consulted.
\subsubsection*{Strings comparison}

One of the challenges in building a co-authorship network is extracting data from a source and correctly assigning the publications to their respective authors. Names may contain errors, for example, names written with different characters, without accents, or the existence of homonyms, that is, two different researchers sharing a similar or equal name that could be identified as the same researcher, thus leading to unreliable data \cite{Savic2019}. To overcome this issue, in this paper, we use a Levenshtein string distance \cite{Levenshtein_SPD66}\cite{doan2012principles} comparison between names to remove or reduce the number of ambiguities, a method extensively used in the literature and which obtained good results in the works of Digiampetri et al.\cite{digiampietri2014}\cite{Digiampietri2012} with data similar to those used in this paper. Explaining this algorithm briefly, it parses 2 strings and returns the number of operations required to transform String A into a String B. If the number of operations is smaller or equal to 2, it is understood that the strings match. Then the combination of the data is made.

\subsubsection*{Generation and definition of networks}
We generate networks of the researchers after the population of the database. We analyzed the programs and their respective researchers in terms of their academic collaborations. In this case, researchers are represented by nodes, and papers published in collaboration between two given researchers represent an edge in the graph. Thus, the higher the ``intraprogram'' collaboration, the higher the collaboration of researchers belonging to the same program, the higher the number of network edges.

By trivially performing a visual analysis, it is possible to perceive the size of the program taking into account the number of researchers in it, since programs with more researchers will have graphs with more nodes than programs with fewer researchers. This visual analysis can also help to visualize the growth of the number of researchers of a program over time because if this number has increased, the graphs will have more nodes than those of previous periods. 

\subsubsection*{Definition and generation of features}
To define the features, we applied graphs algorithms to the networks, and these algorithms return measurements based on the topological analysis of these networks. We arranged these measurements in an array with columns of size \textit{n + 1} where \textit{n} is the total of features, and that total adds CAPES grade in the last column referring to the analyzed program and in the observed period. 

The definition of the features matrix (M\textsubscript{f}) is expressed below, where the columns are varying according to the features going from features \textit{1} (F\textsubscript{1}) to features \textit{n} (F\textsubscript{n}) and in the last column \textit{CAPES grade} of each graduate program, the number of rows $m$ of the matrix varies according to the number of samples since each row is a program in a given CAPES evaluation period.

$$
M\textsubscript{f} = \left[
  \begin{array}{cccccc}
  F\textsubscript{1,1} & F\textsubscript{2,1} & F\textsubscript{3,1} & ... & F\textsubscript{n,1} & NC\textsubscript{1} \\
  F\textsubscript{1,2} & F\textsubscript{2,2} & F\textsubscript{3,2} & ... & F\textsubscript{n,2} & NC\textsubscript{2} \\
  F\textsubscript{1,m} & F\textsubscript{2,m} & F\textsubscript{3,m} & ... & F\textsubscript{n,m} & NC\textsubscript{m} \\
  \end{array}
\right]
$$

\begin{figure}[h!]
\centering 
\includegraphics[width=0.95\textwidth]{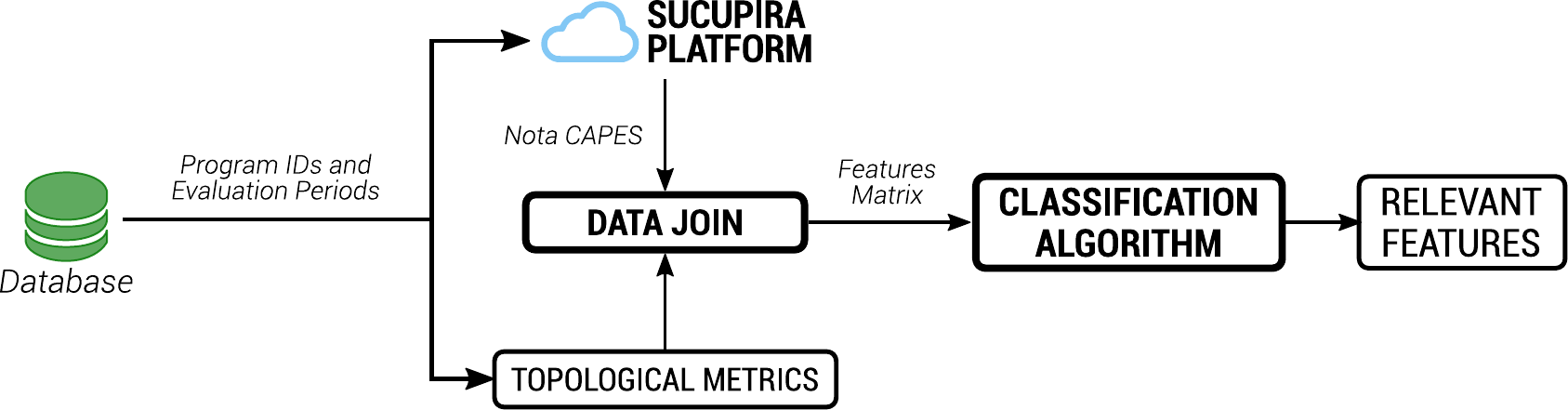}
\caption{\csentence{Process Flow for Topological Metrics Analysis.}
Note that the data saved in the database is used as a filter for downloading CAPES grade from the internet and identifies each program with its topological measurements. This data is gathered into a feature matrix and the algorithms performed.}
\label{fig:Process Flow Topological Metrics}
\end{figure}

The overview of this process are presented in Fig.~\ref{fig:Process Flow Topological Metrics}.

\subsubsection*{Classification and features selection}
We apply classification algorithms on the characteristics matrix, using the data samples from the matrix to classify them in groups where the CAPES score represents each class. It was adopted the  Weka framework \cite{weka2009} and the Random Forest algorithm\cite{Ho1995}, which is a decision-tree algorithm and makes it possible to recover the features used in the classification process. The 10 fold cross-validation was adopted as the validation method as described in \cite{Duda2012}. The choice of this algorithm and its parameters, we explain below.


When executing the algorithm on the matrix, it is possible to recover the importance of each feature in the classification; with this, it is possible to observe which feature has relevance for the correct classification.

To find the best algorithm for the data set, it is also used AutoML tools, which is an automated way executes different feature selection algorithms along with classification algorithms in order to classify the data with higher rate accuracy and lowest error rate. As previously stated, the best algorithm presented for this data set was Random Forest.

\subsubsection*{Author order}
The order in which authors are listed in a publication can provide information that can be analyzed and discussed. This order follows different patterns, depending on the research field or country. A possible way is the alphabetical sorting of the authors; however, this is not typical behavior in the area of Computer Science in Brazil, as well as in other research areas \cite{Costa1992}, in them, authors are often listed according to their contribution \cite{Biagioli2003}\cite{Venkatraman2010}. The first author is the one that used a more considerable effort, defined the materials, methods, and objectives of it and realized the final analysis of the results; already the last author quoted commonly is the supervisor of the work and/or project; the authors who are not in the 2 previous classifications are co-authors who contributed to the work but in a specific way in certain points. The analyzes of these citation orders can be used as qualitative measures of a group of authors' publications.

In this paper, we proposed 3 qualitative indexes to evaluate the order of citation of the authors. The indexes are the \textit{First Author Index}, which is the proportion of all the works in which the authors of the program are the first; the \textit{Contribution Index}, which is the proportion of works in which the researcher is cited at the middle (neither first nor last) and the \textit{Seniority Index}, which is proportional of works in which the researchers are the last author in a paper. The proposal is to compare these indices with the CAPES grade and analyze its possible relation with the proposed indices.

\section*{Results and discussion}

The first step was a pre-processing, applied normalization throughout the data to adjust each adopted feature range from 0 to 1. The analysis algorithms used in the later stages did not suffer interference by the range of values.

Another issue is the unbalanced dataset with 171 graduate programs in which 75 programs with grade 3, 58 programs with grade 4, 14 programs with grade 5, 9 programs with grade 6, and 15 programs with grade 7. Thus, the different number of samples can influence the classification models, since only the programs with grade 3 represented 44\% of the total samples. In this context, we considered subsets of 15 graduate programs with grades 3 and 4 were selected alphabetically by program name. We performed all experiments for each subgroup, and the results considered were the average of these subsets. It is important to note that the figures above refer to the sum of the 3 evaluated periods. Hence, the value exceeds the total of 62 programs being assessed since the same program is repeated up to 3 times, once for each evaluation period.

\begin{figure}[h!]
\centering 
\includegraphics[width=0.95\textwidth]{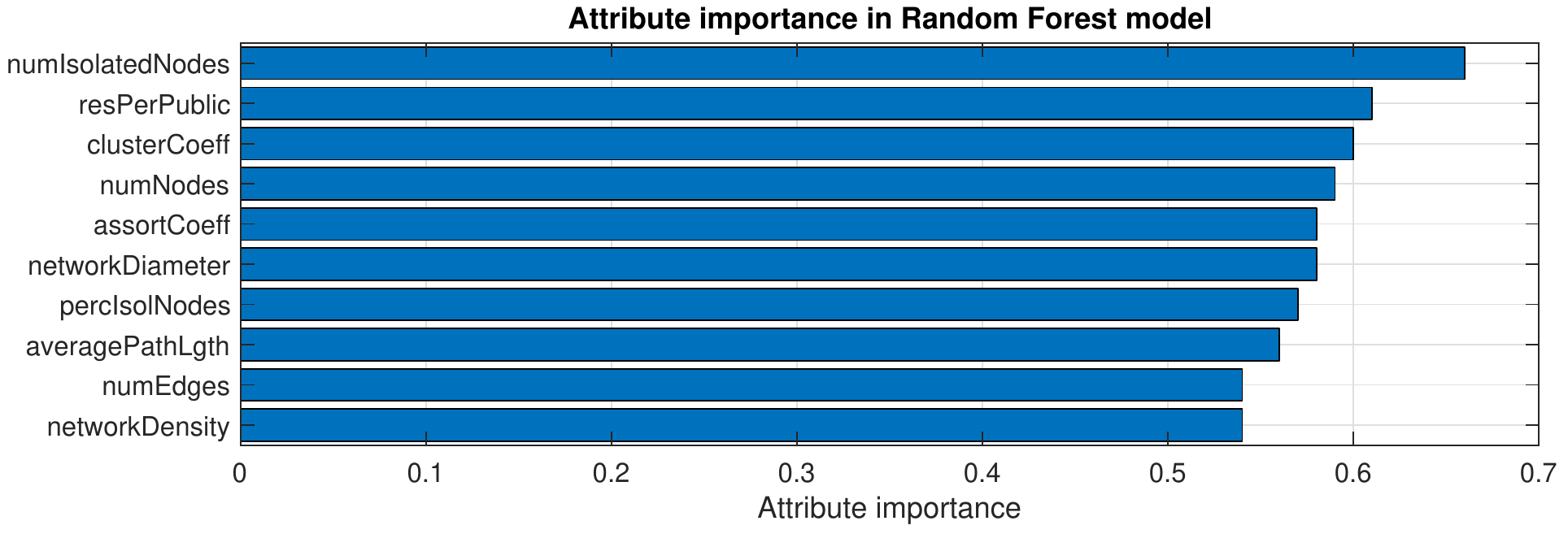}
\caption{\csentence{Importance of each feature for the definition of CAPES grade on Random Forest.} Each of the bars corresponds to a feature, and the larger this bar, the greater the importance of the feature for the Random Forest.}
\label{fig:importance of each feature for the definition of Nota Capes in RF}
\end{figure}

We use the AutoML approach, and on the data of the characteristics matrix, we execute 30 of the supervised classification algorithms implemented by WEKA\cite{weka2009}. The algorithm that obtained the best results was Random Forest, with the test configuration to perform 10 fold cross-validation. Fig.~\ref{fig:importance of each feature for the definition of Nota Capes in RF} shows the percentage of times each feature was selected. It is possible to observe that the most relevant feature in this context is the number of isolated nodes, researchers without any work in common with other researchers.

\begin{figure}[h!]
\centering 
\includegraphics[width=0.95\textwidth]{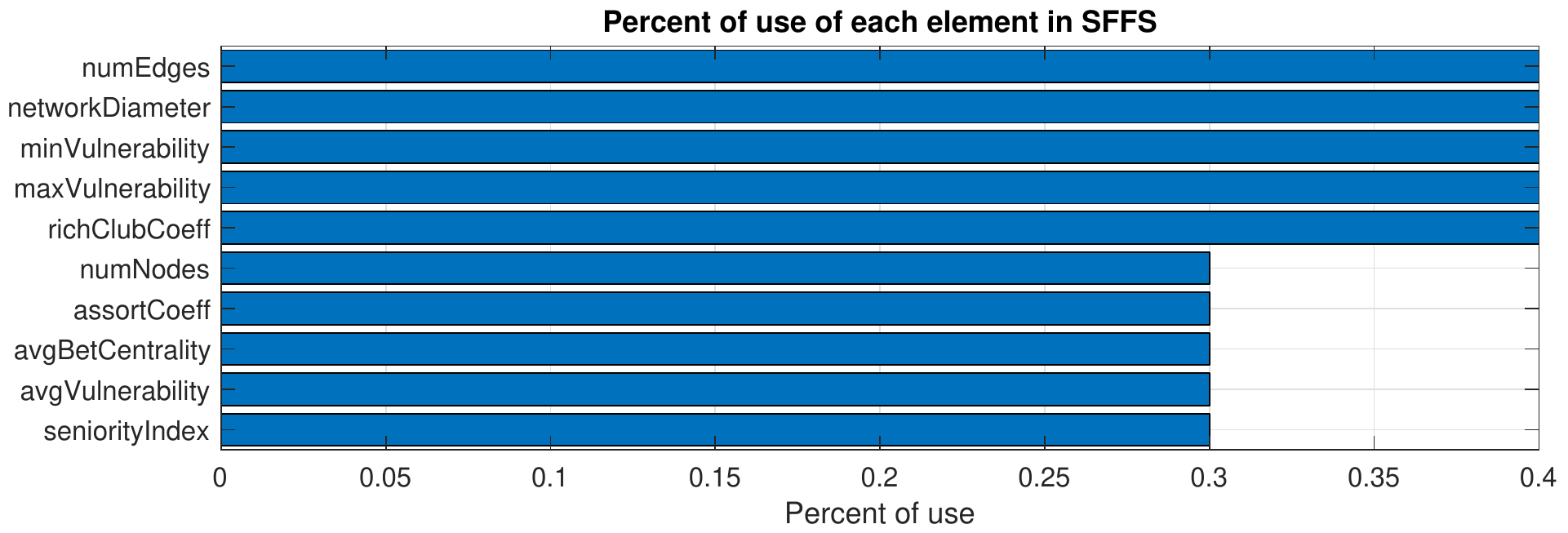}
\caption{\csentence{Importance of each feature for the definition of CAPES grade on SFFS model.} 
Each of the bars corresponds to a feature, and the larger this bar, the more times the feature was used for the SFFS model to classify.}
\label{fig:importance of each feature SFFS}
\end{figure}

We analyzed the feature matrix with the network measurements by a feature selection approach based on the Sequential Forward Floating Selection (SFFS) and the Mean Conditional Entropy as criterion function \cite{lopes2008b,lopes2014a}. For this scheme, just like in the Random Forest, we adopted cross-validation; of the 42 measures used, 8 of them had higher highlights, as represented in the histogram present in Fig.~\ref{fig:importance of each feature SFFS}, which presents the frequency of use of each measurement for the classification of the data set.

\begin{figure}[h!]
\centering 
\includegraphics[width=0.95\textwidth]{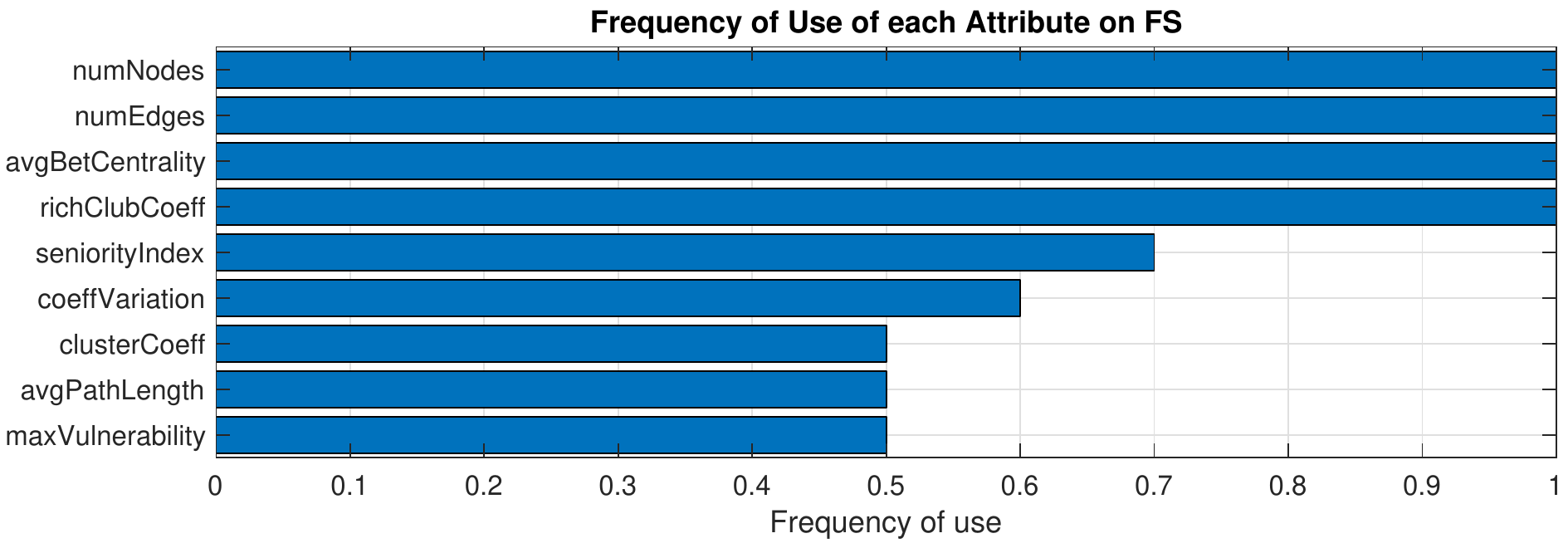}
\caption{\csentence{Importance of each feature for the definition of CAPES grade.}
As in the previous graphs, each bar corresponds to a feature, and the larger this bar, the more often the feature was used for the sort feature selection model.}
\label{fig:importance of each feature for the definition of Nota Capes}
\end{figure}

Using the topological measurements and the 5 classes of CAPES grades, we applied feature selection algorithms to understand which attribute or set had the highest correlation and explain the CAPES grade. The results showed that several features have a high association with the CAPES grade. However, 9 of them had the prominence and a much more significant correlation than the others. We adopted the CfsSubSetEval algorithm as a feature evaluator and BestFirst as the search method for this analysis. The essential features in this analysis are shown in Fig.~\ref{fig:importance of each feature for the definition of Nota Capes} with the percentage of times each feature was selected, and their respective interpretations are explained below.

\begin{itemize}
\item \textit{Number of Nodes:} This measure deals with the number of nodes in each network, in this case, the number of researchers in each program, programs with the highest grade is the programs with the largest number of researchers, so this measure is consistent with the analysis, where the higher the number of nodes/researchers, the greater the CAPES grade; 

\item \textit{Number of Edges:} This measure refers to the number of connections between the network; this context shows the number of publications among the network researchers. Programs with larger CAPES grade are programs that have more works published in collaboration with their researchers. Therefore, it can be stated from the data that the best-evaluated programs have greater internal collaboration compared to the others; 

\item \textit{Average Betweenness Centrality:} This index deals with the nodes' centrality in a network because it analyzes the nodes in the shortest path between 2 connected nodes. The higher the average of this index, the more nodes are participating in shorter routes, so this measures how vital each node to the network is by observing the flow of information that travels through it. In the context of this work, an author with a higher Betweenness Centrality represents the greater direct or indirect participation of him in the publications of the program, with these individual indexes calculates the Average of Betweenness Centrality of the network that in this context indicates the average of partnerships or co-authorship within the program;

\item \textit{Cluster Coefficient:} This index shows the degree of tendency in which the graph nodes have to group and form subsets. In social networks, these clusters are communities of individuals that share common features. In this work context, a cluster is a group of researchers that have research projects in common, so the higher the value of this metric, the greater the number of publications among the researchers of this program (internal collaboration). Programs with upper CAPES grade have greater internal collaboration (higher edges). However, they have a lower cluster coefficient, because although it has individuals with high centrality (importance), these individuals have a lower tendency to form communities; thus, the lowest-rated graduate programs have the feature most similar to small-world networks \cite{Watts1998};

\item \textit{Average Path Length:} This measure performs the average path of the network as the average number of steps in the shortest paths for all possible nodes' possible pairs. The lower this indicator, the higher the efficiency in transporting information from that network. In this work context, this measure represents the average number of authors connecting an Author X to an Author Y, assuming that both do not have a direct connection, so the smaller the measure, the easier it is to connect 2 directly disconnected researchers. The data presented by this work show that programs with lower evaluations in the CAPES grade have a lower value in this measure. Better-evaluated programs have higher complexity in the connection of their nodes. Since the number of nodes metric and the number of edges in better-evaluated programs is higher, there are more paths (co-authorship) and more nodes (researchers) in these networks; however, when this metric is considered relatively, that is, divide the value of the Average Path of the Network by the number of edges it is possible to notice that this value is inverted; therefore, it indicates that although the networks are bigger, they are still efficient;

\item \textit{Rich Club Coefficient:} This index measures the proportion in which the nodes with a significant number of connections (hubs) of a network are connected. This coefficient can be used to evaluate the robustness of a network, because the higher the value, the more strongly connected, which indicates that if one of these hubs is removed, the lower the impact on the network structure. In the context of this work, this measure evaluates the tolerance to changes of a program if a researcher is randomly removed, programs with lower CAPES grade have a greater dependence on their researchers, in case a vital researcher is removed from the program, the structure of the program will be strongly affected, different from the best-evaluated programs, that have less individual dependence of researchers;

\item \textit{Variation Coefficient:} The Coefficient of Variation measures the variation of a network; its mathematical equation is the result of the standard deviation of the values of the network divided by the average of these values; in the approach of this project, this metric informs how much the measures vary in each one of these programs;

\item \textit{Swan Connectivity:} The Swan Connectivity function is a measure of network vulnerability \cite{Lhomme2015}, which calculates the loss of connectivity when a vertex is removed from the network; the result of this calculation measures the decrease in the number of relationships between each vertex of the network when one vertex or several are removed. In this project's context, this measure indicates how vulnerable a network can be because when removing a researcher (vertex), the network tends to lose connectivity.

\item \textit{Seniority Index:} This measure is proposed in order to evaluate the percentage of publications in which the researcher has the last name in a publication; thus, it is possible to qualify the publications of each researcher and therefore generalize this measure as the average of the Seniority Index of all researchers in the same graduate program.

\end{itemize}

A third adopted approach was to apply the Spearman Correlation algorithm \cite{Spearman1904} on the feature matrix that returned the data consistent with the classification measures, and the three features with the highest correlation with the CAPES grade remained the same. Table~\ref{tab:metric-correlation}1 shows the ten measurements more relevant and their respective associations with the CAPES grade. The left column is informed of which measure is used. In the right, the correlation percentage and when the measure has values greater than 0 (positive) means that it is directly correlated with the CAPES grade, but if the value is less than 0 (negative) represents that this measure is inversely proportional so when the CAPES grade increases, the measurement decreases.

\begin{table}[h!]
\label{tab:metric-correlation}
\caption{Measurements correlation with CAPES grade.}
\begin{tabular}{@{}lr@{}}
\toprule
\textbf{Metric of Matrix}         & \textbf{Correlation}  \\ \midrule
Number of Nodes                   & 0.638  \\ 
Minimum Swan Connectivity         & -0,633  \\
Number of Edges                   & 0.632  \\
Netwrork Diameter                 & 0.595  \\
Average Path Length               & 0.553  \\
Max Swan Connectivity             & 0.505  \\
Eigen Centrality                  & 0.473  \\
Average Degree Centrality         & 0.440 \\
Average Path Length               & 0.440  \\
Average Betweenness Centrality    & 0.096  \\
\bottomrule
\end{tabular}
\end{table}

To better understand the data set, some classifications by decreasing the number of the class were performed. The initial 5 class, which were CAPES grade 3, 4, 5, 6 and 7, were grouped into 3 categories: grades 3 and 4 in a group Programs C, programs with CAPES grade 5 became Programs B, and programs with CAPES grades 6 and 7 were grouped as Programs A. 
When analyzing the programs clustered in this way, it can be seen that Programs A differ significantly from Programs C in relation to topological measurements, this intensifies the importance of analyzing these measures. With a lower number of classes, the algorithms perform more accurately, achieving 96.47\% accuracy rate. 

\subsubsection*{Proposed Indexes}
When analyzing them, the proposed indexes can be verified that the Seniority Index has a high correlation with the CAPES grade.
Therefore, based on the results, it can be indicated that the best-evaluated programs have a higher number of works where the members of the programs are cited as the last authors. Besides, it is possible to notice that programs with lower CAPES grades have a higher first author rate. 

\begin{figure}[h!]
\centering 
\includegraphics[width=0.95\textwidth]{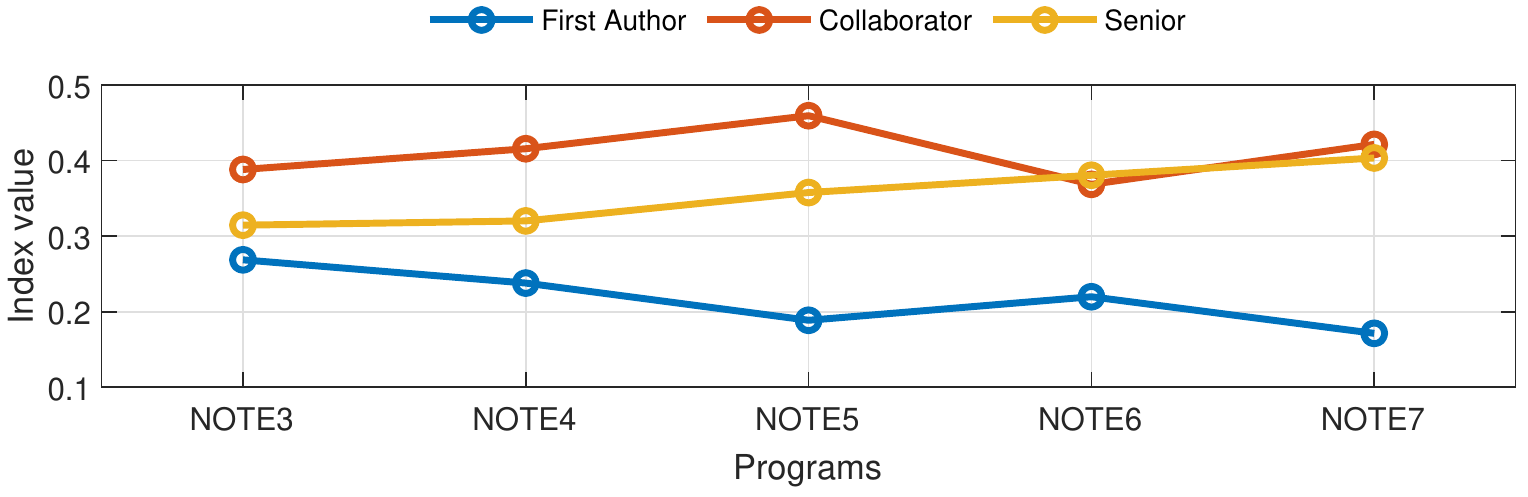}
\caption{\csentence{Average of the proposed indexes in each graduate program by considering different CAPES grade.}
Each row represents one of the three proposed indices; the CAPES grade is on the x-axis.}
\label{fig:average of the proposed indices in each program group}
\end{figure}

Fig~\ref{fig:average of the proposed indices in each program group} shows the average of proposed indexes using the 5 classes. It is possible to observe that the Seniority Index has a linear rise according to the increase in the programs' evaluation. The Seniority Index has its lowest value for programs with grade 3 and its highest value for programs with grade 7. The First Author Index presents the inverse behavior, and its peak is for programs with grade 3, indicating a pattern of research composition on these graduate programs.
It is also possible to observe that the grades 6 and 7 have a very close average between Collaborator and Senior.

\begin{figure}[h!]
\centering 
\includegraphics[width=0.95\textwidth]{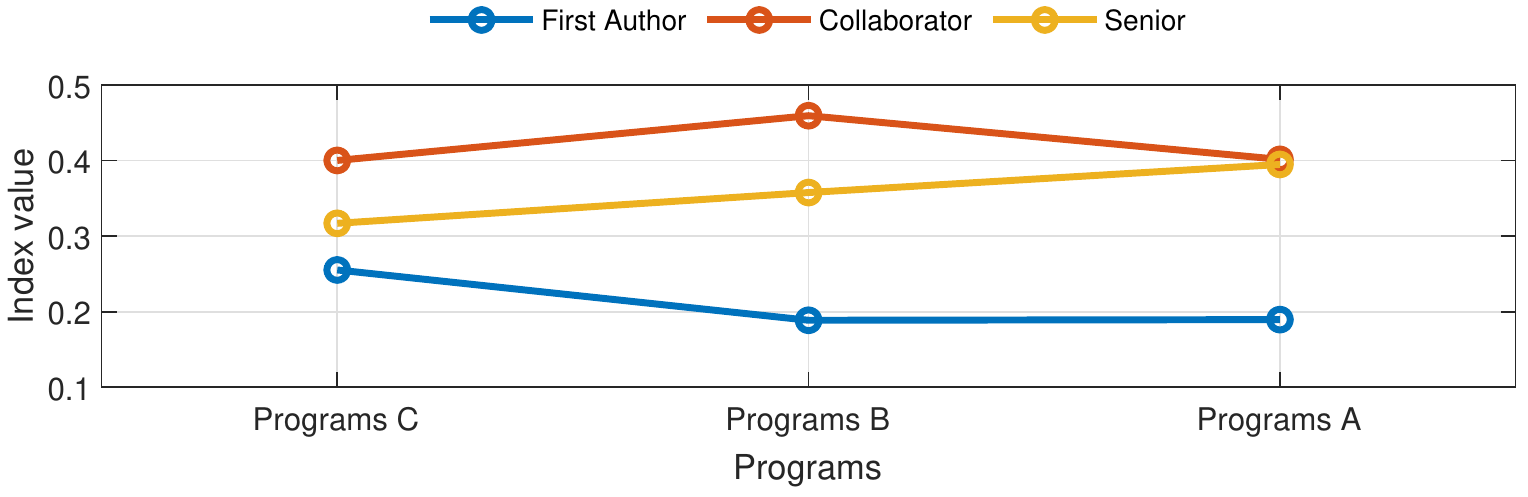}
\caption{\csentence{Average of the proposed indexes in each program group, clusters in 3.}
Each line represents one of the three proposed indexes, but the programs are clustered into 3 and are represented on the x-axis.}
\label{fig:average of the proposed indices in each program group, cluster3}
\end{figure}

Figure~\ref{fig:average of the proposed indices in each program group, cluster3} presents the proposed indexes behavior for the proposed by considering 3 clusters of graduate program grades. It is possible to notice that proposed indexes complement each other since there are a high Contribution and Seniority Index, with a low First Author Index, which makes the measures coherent since they are the averages of the programs. The same analysis starting from the grouping of the programs in only 3 classes (Programs A, ..., Programs C) was performed. In this classification, the curves are less varied, and the trend in the measurements is more evident. The Seniority Index always keeps linear and ascending according to the CAPES grade, the Collaboration Index peaks in intermediate programs, and the First Author Index is decrescent with higher value in lower-grade programs.

\begin{figure}[h!]
\centering 
\includegraphics[width=0.95\textwidth]{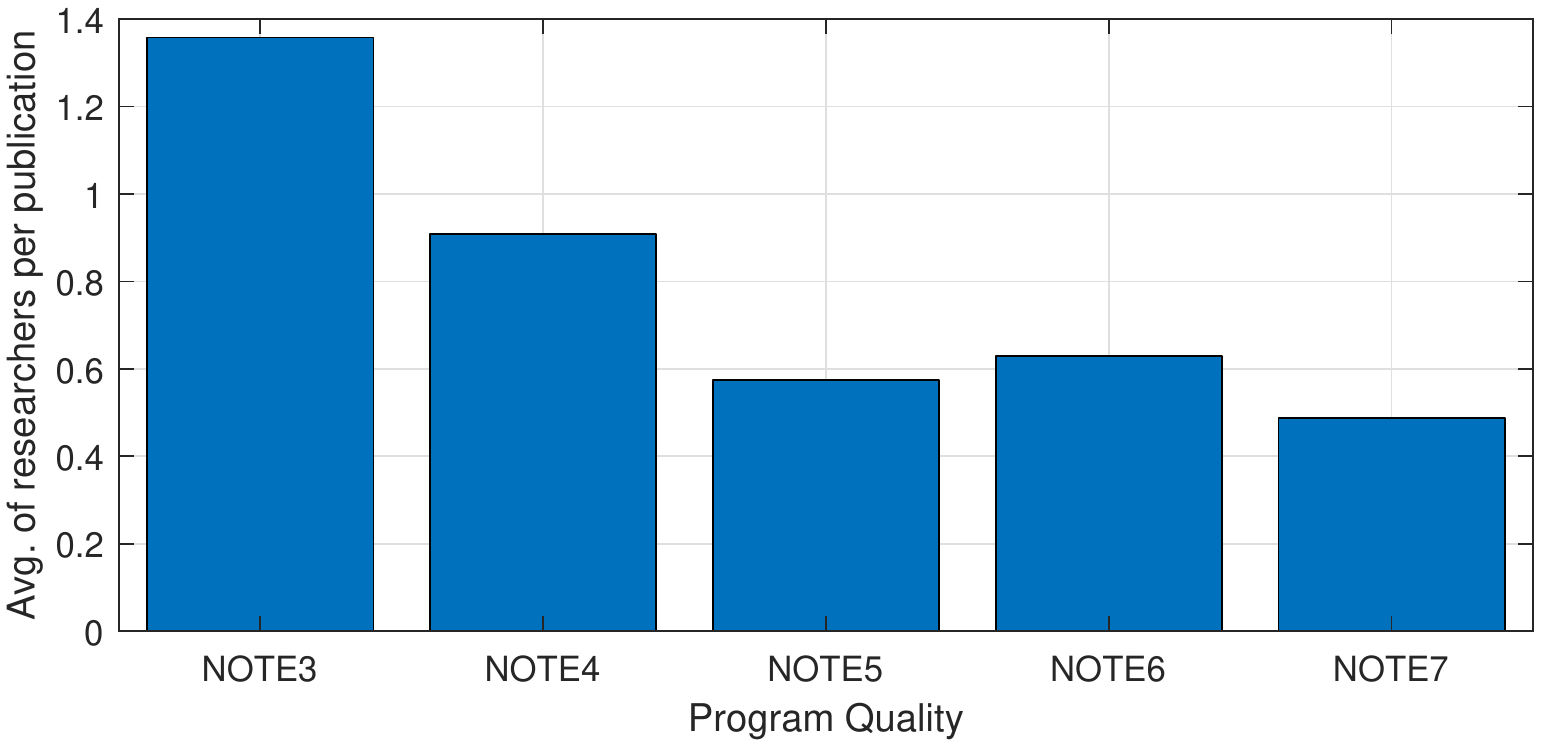}
\caption{\csentence{Average Number of Researchers per Publication X CAPES grade.}
Each bar represents CAPES grades and the higher these bars are, the higher the average number of researchers per publications.}
\label{fig:Average Number of Researchers per Publication X Nota Capes}
\end{figure}

The Average Number of Researchers per Publications, although very simple, can be handy because it is the result of dividing the number of nodes by the number of edges. When analyzing the data, it can be verified that programs with higher CAPES grade have a lower average number of researchers per work. Therefore, it can be stated that these are more efficient than the programs with lower CAPES grades. Fig.~\ref{fig:Average Number of Researchers per Publication X Nota Capes} corroborates with the information, indicating that the average number of researchers per publication for each CAPES grade. It is possible to observe a pattern indicating that programs with lower CAPES grades present a higher average number of researchers per publication. In comparision, programs with higher CAPES grades have a lower average number of researchers per publication.

To strengthen the justification of the importance of each measure, the results of all of the adopted measurements were compared to the CAPES grade, and the final value is presented in Table~\ref{tab:most relevant measure}2. The adopted strategy was as follows: Weka tool was adopted to perform a feature selection and the same in pointed to some measures as having a higher correlation with CAPES grade; the Random Forest classification algorithm was performed, and the same told us the measures as of greater relevance next to CAPES grade; on the data we also applied the Spearman Correlation algorithm that showed us the most correlated measure to the CAPES grade; Finally, the feature selection algorithm (SFFS and the mean conditional entropy, as a multivariate analysis) to observing the most relevant measurements. The results of each approach are different since they are quite different algorithms in their conceptions. However, some measures present highlights because they are treated as of great relevance in more than one approach, according to Table~\ref{tab:most relevant measure}2 that shows the ten most relevant measures.

\begin{table}[h!]
\label{tab:most relevant measure}
\caption{Most relevant measures}
\centering
\begin{tabular}{@{}rcccccc@{}} 
\toprule
\textit{\textbf{Metric/Algorithm}} & \textbf{Sel.Att~} & \textbf{RForest} & \textbf{Spearman~} & \textbf{SFFS} & \textbf{Total} \\ 
\midrule
\textit{numberNodes} & Y & Y & Y & Y & 4 \\
\textit{numberEdges}& Y & Y & Y & Y & 4 \\
\textit{clusterCoefficient}& Y & Y & N & N & 2 \\
\textit{avgPathLength}& Y & Y & Y & N & 3 \\
\textit{avgBetweennessCentrality} & Y & N & Y & Y & 3 \\
\textit{maxVulnerability} & Y & N & N& Y & 2 \\
\textit{richClubCoefficient}& Y & N & N & Y & 2 \\
\textit{coefficientVariation} & Y & N & N & N & 1 \\
\textit{seniorityIndex} & Y & N & N & Y & 2 \\
\textit{numberIsolatedNodes} & N & Y & N & N & 1 \\
\textit{networkDiameter}& N & Y & Y & Y & 3 \\
\textit{assortCoefficient} & N & Y & N & Y & 2 \\
\textit{percIsolatedNodes} & N & Y & N & N & 1 \\
\textit{ResearchersPerPublication}~ & N & Y & N & N & 1 \\
\bottomrule
\end{tabular}
\end{table}

\section*{Conclusions}

This work proposes a complex network approach to analyze and characterize the networks of co-authorships of Brazilian computer science graduate programs. More specifically, the graduate program co-authorship networks were analyzed. We identified the most relevant network measurements to characterize these graduate programs' quality by considering the CAPES grade.

We analyzed 62 Brazilian graduate programs in computer science, with about 1,644 researchers observed in three CAPES evaluation periods from 2007 to 2016. The data were acquired in 2019 to ensure that the researchers' productions were complete for the period evaluated. The obtained results in these analyzes were significant since measuring network analysis measures can indicate the CAPES grade with high accuracy.

The adopted measurements that considering the size of the networks (programs) were the most significant, and several learning models recognize them as having a high correlation with the CAPES assessment. The larger the program, either in the number of researchers or in the number of publications, Higher is your evaluation scores. We understand that these measures are trivial, and if they were to be taken into account only them, it would not be necessary to model the way we did in this work; however, the objective of this work is not only to point out these measures but a series of measures that when grouped can deliver significant results. Although the measures are trivial, we managed to prove that they are important through network analysis.

The programs also point to the measures of centrality (importance) of some researchers in the network, and we observed some standards. We observed that better-evaluated programs have more elements with higher centrality, so these programs have more researchers of greater influence. Vulnerability (Rich Club Coefficient and SWAN Connectivity) measures also yielded relevant results, with better-rated programs being less vulnerable than less-rated ones. When a researcher is randomly removed, the program structure undergoes fewer changes, which is not the case with less-rated programs that are more highly rated vulnerable. If a researcher is removed, the structure of that program will undergo significant changes.

We proposed a measure of quantitative analysis of the average of researchers that publications have in the programs; this proposed measure delivered relevant results where it was possible to observe that programs but well evaluated, are more productive because they have more publications for each researcher, therefore, lowest Average Researchers by Publications.

We proposed three qualitative measures of collaborative evaluation among the researchers, and the pattern that was observed is that programs with lower CAPES grade have a higher First Author Index than the others; Intermediate CAPES grade programs have a higher Collaboration Index than others; and lastly, the highest-rated CAPES programs have a higher Seniority Index than the others, and this index has a linear behavior that causes it to grow linearly as the program evaluation increases.

As a result, this work points out some important patterns to be analyzed that lead to the characterization of the graduate programs related to CAPES grade that can bring information for the Brazillian computer science community to analyze and to adopt strategies that can lead to the improvement of these patterns and, consequently, improve the evaluation of the graduate programs.

\section*{List of abbreviations}
The abbreviations used in this paper and their respective meanings are presented below:\\

\newcommand{\abbrlabel}[1]{\makebox[3.2cm][l]{{#1}\ \dotfill}}
\newenvironment{abbreviations}{\begin{list}{}{\renewcommand{\makelabel}{\abbrlabel}}}{\end{list}}

\begin{abbreviations}
    \item [AutoML]	Automated Machine Learning
    \item [CAPES]	\textit{Coordenação de Aperfeiçoamento de Pessoal de Nível Superior}
    \item [CV]	\textit{Curriculum Vitae}
    \item [DBLP]	Digital Bibliography \& Library Project
    \item [DBMS]	Database Management Systems
    \item [FS]	Feature Selection
    \item [JCR]	Journal Citation Reports
    \item [ORCID]	Open Researcher and Contributor ID
    \item [SFFS]	Sequential Forward Floating Selection
    \item [SJR]	Scientific Journal Rankings
    \item [XML]	eXtensible Markup Language

\end{abbreviations}

\begin{backmatter}

\section*{Availability of supporting data}
The datasets generated and/or analysed during the current study are available in the Github repository, \url{https://github.com/alexjrns/datamining_lattes_computer_science}

\section*{Competing interests}
The authors declare that they have no competing interests.

\section*{Funding}
Conselho Nacional de Desenvolvimento Científico e Tecnológico (CNPq) [406099/2016-2]. Fundação Araucária e do Governo do Estado do Paraná/SETI. Federal University of Technology – Paraná.

\section*{Author's contributions}
AJNS and MMB wrote the software code, the data set, analyzed the data and wrote the manuscript. FML and JPMC participated in the design, analyzed the data and coordination of the work. All authors contributed to, read, revise and approve the final manuscript.

\section*{Acknowledgements}
Not applicable.
  

\bibliographystyle{bmc-mathphys} 
\bibliography{references.bib}  

\section*{List of Figures}

\textbf{Figure~\ref{fig:Process Flow2} Process Flow.} Clouds indicate internet data access. Blocks in bold represent processes. Artifacts generated by each process are described next to the arrows that indicate the directions of the flow.

\textbf{Figure~\ref{fig:Process Flow Topological Metrics} Process Flow for Topological Metrics Analysis.}
Note that the data saved in the database is used as a filter for downloading CAPES grade from the internet and also serves as identification of each program with its topological measurements, this data is gathered into a matrix of characteristics where the algorithms are applied.

\textbf{Figure~\ref{fig:importance of each feature for the definition of Nota Capes in RF} Importance of each feature for the definition of CAPES Grade on Random Forest.} Each of the bars corresponds to a characteristic and the larger this bar, the greater the importance of the attribute for the Random Forest model to rank.

\textbf{Figure~\ref{fig:importance of each feature SFFS} Importance of each feature for the definition of CAPES Grade on feature selection approach.} 
Each of the bars corresponds to a characteristic and the larger this bar, the more times the attribute was used for the SFFS model to classify.

\textbf{Figure~\ref{fig:importance of each feature for the definition of Nota Capes} Importance of each feature for the definition of CAPES Grade.}
As in the previous graphs, each bar corresponds to a characteristic and the larger this bar, the more often the attribute was used for the sort attribute selection model.

\textbf{Figure~\ref{fig:average of the proposed indices in each program group} Average of the proposed indexes in each program group.}
Each row represents one of the three proposed indices; the 7 program classification groups are represented on the x axis.

\textbf{Figure~\ref{fig:average of the proposed indices in each program group, cluster3} Average of the proposed indexes in each program group, clusters in 3.}
Each line represents one of the three proposed indexes but the program classification groups are divided into 3 and are represented on the x axis.

\textbf{Figure~\ref{fig:Average Number of Researchers per Publication X Nota Capes} Average Number of Researchers per Publication X CAPES Grade}
Each bar represents one of seven program groups, and the larger these bars are, the higher the average number of researchers per program post.

\end{backmatter}
\end{document}